\documentclass{article}
\usepackage{multirow}
\usepackage{cite}
\usepackage{url}

\usepackage{graphicx}
\usepackage{subcaption}
\usepackage{amsmath}
\usepackage[psamsfonts]{amssymb}
\usepackage{amsxtra,spconf}
\usepackage{threeparttable}

\usepackage{adjustbox,tabularx}
\usepackage{float}

\name{Yin-Ping Cho\textsuperscript{1}, Yu Tsao\textsuperscript{2}, Hsin-Min Wang\textsuperscript{3}, Yi-Wen Liu\textsuperscript{1}}
\address{\textsuperscript{1} National Tsing Hua University, Hsinchu, Taiwan\\
	\textsuperscript{2} Research Center for Information Technology Innovation, Academia Sinica, Taipei, Taiwan\\
	\textsuperscript{3} Institute of Information Science, Academia Sinica, Taipei, Taiwan
}

\title{Mandarin Singing Voice Synthesis with Denoising Diffusion Probabilistic Wasserstein GAN}

\begin{document}
	

	\maketitle
	
	\begin{abstract}
		Singing voice synthesis (SVS) is the computer production of a human-like singing voice from given musical scores. To accomplish end-to-end SVS effectively and efficiently, this work adopts the acoustic model-neural vocoder architecture established for high-quality speech and singing voice synthesis. Specifically, this work aims to pursue a higher level of expressiveness in synthesized voices by combining the diffusion denoising probabilistic model (DDPM) and \emph{Wasserstein} generative adversarial network (WGAN) to construct the backbone of the acoustic model. On top of the proposed acoustic model, a HiFi-GAN neural vocoder is adopted with integrated fine-tuning to ensure optimal synthesis quality for the resulting end-to-end SVS system. This end-to-end system was evaluated with the multi-singer Mpop600 Mandarin singing voice dataset. In the experiments, the proposed system exhibits improvements over previous landmark counterparts in terms of musical expressiveness and high-frequency acoustic details. Moreover, the adversarial acoustic model converged stably without the need to enforce reconstruction objectives, indicating the convergence stability of the proposed DDPM and WGAN combined architecture over alternative GAN-based SVS systems.\footnote{Evaluation audio samples can be found at: \url{https://yinping-cho.github.io/diffwgansvs.github.io/}}
		
	\end{abstract}
	
	\section{Introduction}
	Singing voice synthesis (SVS) aims to generate singing voices as natural and expressive as those of human singers. The attention from both academia and commercial corporations has pushed the boundaries of SVS with neural networks in recent years.
	
	Although there are neural network-driven SVS systems designed to generate waveforms directly from the musical scores \cite{wavenet, VIsinger}, the most prominent system design is to split the generation pipeline into two stages. First, an \emph{acoustic model} frontend consumes the musical score to estimate the intermediate acoustic features for the singing voice; then, a backend synthesizes the final audio waveform from those acoustic features. While the intermediate acoustic features can be spectrograms \cite{korean_gansinger} or the \emph{WORLD} vocoder \cite{world} parameters \cite{xiaoicesing, litesing}, Mel-spectrograms are preferred by most neural SVS systems pursuing the highest audio quality as in \cite{fftsinger, hifisinger, diffsinger}. These Mel-spectrogram intermediates are transformed to audio waveforms with neural vocoder backends such as  Parallel WaveGAN \cite{pwg}, HiFi-GAN \cite{hifigan}, or the novel singing-specific SawSing \cite{sawsing} neural vocoder.
	
	In this setup, the acoustic model determines how musically natural and expressive the synthesized singing voice will be. Numerous deep learning techniques have been deployed as backbone modules for constructing acoustic models. Some of the noticeable techniques in the landmark systems include long-short term memory machine (LSTM) \cite{lstm_korean, lstm_mandarin}, WaveNet \cite{wavenet, litesing}, and Transformer \cite{xiaoicesing, fftsinger}. These systems commonly employ reconstructive L1 or L2 loss to train the acoustic model to estimate the acoustic features. Nevertheless, generative models trained on simple reconstructive targets often suffer from over-smoothing and producing estimations that approach the mean/median of the target distribution but lack human-like variations.
	
	Therefore, some of the recent advancements in SVS acoustic models are based on generative adversarial networks (GANs) and denoising diffusion probabilistic models (DDPMs). Regarding musical naturalness and expressiveness, GANs have proven to be an enhancement in multiple SVS systems \cite{korean_gansinger, japanese_gansinger, hifisinger}, but they are usually less stable in training and may require dataset-specific hyperparameters. On the other hand, DDPM is a generative model that yield promising performance across domains, such as image generation \cite{ddpm, ddpm_beats_gan}, neural vocoding \cite{diffwave}, speech enhancement \cite{lu2022conditional}, and speech or singing voice synthesis \cite{difftts, diffsinger}. Considering their high performance, the confluence of the two techniques was a natural development. The combination of GAN and DDPM has shown excellent performances in image generation \cite{diffgan} with a subsequent attempt in speech synthesis \cite{diffgan_tts}.
	
	To explore such a combined architecture of DDPM and GAN, this work trains the DDPM acoustic model adversarially with a \emph{Wasserstein} GAN (WGAN) algorithm. Among different kinds of GANs, WGAN is the one that promises to approximate the real data distribution even when this probability density may not be approachable by other GAN metrics \cite{wgan, wgan_div}. This property makes it an ideal GAN algorithm for SVS where the complicated singing voice distribution is usually supported by a relatively small dataset, owing to data collection diffculties. Additionally, this work proposes a \emph{Musical-Score-Conditioned} (MSC)-discriminator that incorporates information of the sung contents to further prevent mode collapses that are often encountered in GAN-based synthesis models.
	
	\section{Denoising Diffusion Wasserstein GAN}
	The following section introduces DDPM with \emph{Wasserstein GAN} (WGAN) \cite{wgan}. Section ~\ref{sec:ddpm} states the denoising diffusion probabilistic model formulation, Section ~\ref{sec:diffwgan} explores why and how a Wasserstein GAN is applied to DDPM, and Section ~\ref{sec:diffwgan-svs} presents the denoising diffusion WGAN formulated for SVS in this work.
	
	\subsection{Diffusion Denoising Probabilistic Model} \label{sec:ddpm}
\begin{figure}[H]
		\centering
		\includegraphics[width=0.99\linewidth]{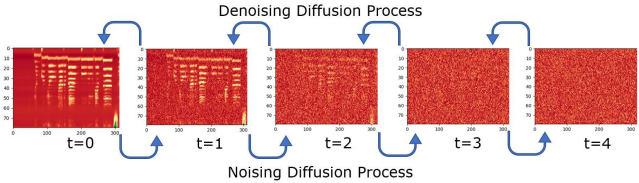}
		\caption{Visualization of the \emph{noising diffusion process} and the \emph{denoising diffusion process} with a Mel-spectrogram.}
		\label{fig:diff_process}
	\end{figure}

	The working of DDPM is characterized by a Gaussian Markov chain with a \emph{ noising process} and a \emph{denoising process}, which are visualized in Fig. ~\ref{fig:diff_process}.
	
	The \emph{noising process} transforms a true data sample $\mathbf{x}_0$, in this case a clean Mel-spectrogram, to a unit normal noise $\mathbf{x}_T \sim \mathcal{N}(\mathbf{0},\mathbf{I})$ of the same dimensions, formulated as:
	\begin{equation} \label{noising}
		\begin{gathered}
			q( \mathbf{x}_{1:T}|\mathbf{x}_0 ) = \prod_{\substack{t \geq 1}} q( \mathbf{x}_t|\mathbf{x}_{t-1} ),  t=1, 2, 3,...,T;\\
			q( \mathbf{x}_t|\mathbf{x}_{t-1} ) = \mathcal{N}( \mathbf{x}_t; \sqrt{1-\beta_t}\mathbf{x}_{t-1}, \beta_t\mathbf{I} ),\\
		\end{gathered}
	\end{equation}
	where the variance schedule $\{\beta_t$\} is pre-defined according to an exponential scheme same as in \cite{diffgan}. Conversely, $q(\mathbf{x}_0)$ denotes the data generation of a true Mel-spectrogram, and $q(\mathbf{x}_{t-1}|\mathbf{x}_t)$ is thus the true denoising diffusion transition.
	
	The \emph{denoising process} gradually synthesizes an estimated Mel-spectrogram $\mathbf{\hat{x}}_0$ from a random initial noise $\mathbf{x}_T$ with model parameter $\theta$ in the form of a \emph{denoising} process:
	\begin{equation} \label{denoising}
		\begin{gathered}
			p_{\theta}( \mathbf{\hat{x}}_{0:T} ) = p(\mathbf{x}_T)\prod_{\substack{t \geq 1}} p_{\theta}( \mathbf{\hat{x}}_{t-1}|\mathbf{x}_t ), t=T,T-1,...2,1;\\
			p_{\theta}( \mathbf{\hat{x}}_{t-1}|\mathbf{x}_t ) = \mathcal{N}( \mathbf{\hat{x}}_{t-1}; \mathbf{\mu}_\theta(\mathbf{x}_t,t), \sigma_t^2\mathbf{I} ),
		\end{gathered}
	\end{equation}
	where $\sigma_t^2$ is fixed according to each $t$, and $\mathbf{\mu}_\theta(\mathbf{x}_t,t)$ is estimated with conditions $\mathbf{x}_t$, $t$, and model parameter $\theta$.
	
	Here, same as in \cite{diffgan}, we choose to parameterize the denoising transition by first estimating a clean Mel-spectrogram $\mathbf{\hat{x}}_{0,t}$ directly from noised data $\mathbf{x}_t$ and use it to sample the estimated denoised  $\mathbf{\hat{x}}_{t-1}$:
	\begin{equation} \label{est_x_t-1}
		p_{\theta}( \mathbf{\hat{x}}_{t-1}|\mathbf{x}_t ):=q( \mathbf{\hat{x}}_{t-1}|\mathbf{x}_t, \mathbf{\hat{x}}_{0,t}=f_\theta(\mathbf{x}_t, t) ),
	\end{equation}
	where the denoising distribution $f_\theta(\mathbf{x}_t, t)$ is estimated by the acoustic model generation $f_\theta(\mathbf{x}_t, t)=G_{\theta}(\mathbf{x}_t, t, \mathbf{ms}, id)$, with $\mathbf{ms}$ denoting the \emph{musical score} with lyrics, \emph{id} being the \emph{singer identity}, and $G_{\theta}(\cdot)$ as the neural network generator with learnable parameter $\theta$.
	
	Since the diffusion sampling is simply a Gaussian sampling, $\mathbf{\hat{x}}_t$ can be computed in one step as:
	\begin{equation} \label{x_t}
		\begin{gathered}
			\mathbf{\hat{x}}_t = \sqrt{\bar{\alpha}_{t}}\mathbf{\hat{x}}_0 + \sqrt{1-\bar{\alpha}_{t}} \mathbf{\epsilon},
			\mathbf{\epsilon} \sim \mathcal{N}(\mathbf{0},\mathbf{I})
		\end{gathered}
	\end{equation}
	and $q( \mathbf{\hat{x}}_{t-1}|\mathbf{x}_t, \mathbf{\hat{x}}_0)$ is efficiently obtained with
	\begin{equation} \label{x_t-1}
		\begin{gathered}
			q( \mathbf{x}_{t-1}|\mathbf{x}_t, \mathbf{x}_0) = \mathcal{N}(\mathbf{x}_{t-1}; \tilde{\mathbf{\mu}}((\mathbf{x}_0,\mathbf{x}_t),\tilde{\beta}_t\mathbf{I})),\\
			\tilde{\mathbf{\mu}}(\mathbf{x}_0,\mathbf{x}_t) = \frac{\sqrt{\bar{\alpha}_{t-1}}\beta_t}{1-\bar{\alpha}_{t}}\mathbf{x}_0 + \frac{\sqrt{\alpha_t}(1-\bar{\alpha}_{t-1})}{1-\bar{\alpha}_{t}}\mathbf{x}_t
		\end{gathered}
	\end{equation}
	where the scalar constants $\alpha_t:=1-\beta_t$, $\bar{\alpha}_t:=\prod_{s=1}^{t}\alpha_s$, and $\tilde{\beta}:=\frac{1-\bar{\alpha}_{t-1}}{1-\bar{\alpha}_t}\beta_t$ can be pre-computed once the variance schedule $\{\beta_t\}$ is defined as a hyperparameter of the process \cite{ddpm, diffgan}.
	
	\subsection{Diffusion Wasserstein GAN} \label{sec:diffwgan}
	To achieve good synthesis quality with the \emph{denoising process} introduced in Subsection ~\ref{sec:ddpm}, DDPMs usually set the number of time steps $T$ to a large number (hundreds to thousands) and keep ${\beta_t}$ very small. This way, both the \emph{noising} $q( \mathbf{x}_t|\mathbf{x}_{t-1} )$ and the \emph{denoising} $q( \mathbf{x}_{t-1}|\mathbf{x}_t )$ diffusion transitions of the true data are sufficiently Gaussian, making it easy for the model to approximate the true denoising transition: $p_{\theta}( \mathbf{\hat{x}}_{t-1}|\mathbf{x}_t ) \approx q(\mathbf{x}_{t-1}|\mathbf{x}_t)$.
	
	However, setting a large $T$ and small $\{\beta_t\}$ makes the synthesis process much more time-consuming both in the training and the inference phases. Consequently, we adopt a more efficient Variance Preserving (VP) SDE scheme proposed in \cite{variance_schedule}, computing $\{\beta_t\}$ according to
	\begin{equation}
		\beta_{t} = 1-\mathrm{exp} \{ {
			\beta_{\mathrm{min}}(\frac{1}{T}) - 0.5 (\beta_{\mathrm{max}} - \beta_{\mathrm{min}})\frac{2t-1}{T^2}
		} \},
	\end{equation}
	where we set $\beta_{\mathrm{max}}=20$, $\beta_{\mathrm{min}}=0.1$, and the number of time steps $T=4$, found as the best-performing hyperparameters in \cite{diffgan}.
	
	Nevertheless, this few-step, big-jump diffusion scheme poses a much greater challenge to the acoustic model. Since the distribution of Mel-spectrograms is apparently not Gaussian and is empirically sparse, the diffusion transition $q(\mathbf{x}_{t-1}|\mathbf{x}_t)$ is a complicated conditional distribution that is hard to estimate.
	
	Thus, the objective for the diffusion denoising generator is formulated as
	\begin{equation}
		\min_{\substack{\theta}} \sum_{\substack{t \geq 1}}\mathbb{E}_{q(\mathbf{x}_t)}[D_{\mathrm{adv}} (q(\mathbf{x}_{t-1}|\mathbf{x}_t) || p_{\theta}( \mathbf{\hat{x}}_{t-1}|\mathbf{x}_t )) ],
	\end{equation}
	which aims to match the true transition $q(\mathbf{x}_{t-1}|\mathbf{x}_t)$ and the model's synthesis transition $p_{\theta}( \mathbf{\hat{x}}_{t-1}|\mathbf{x}_t )$ by minimizing an \emph{adversarial objective} $D_{adv}$ that estimates the \emph{Wasserstein Distance} with adversarial neural networks.
	
	\begin{figure*}[t]
		\centering
		\begin{subfigure}[t]{.49\textwidth}
			\centering
			\includegraphics[width=0.8\textwidth]{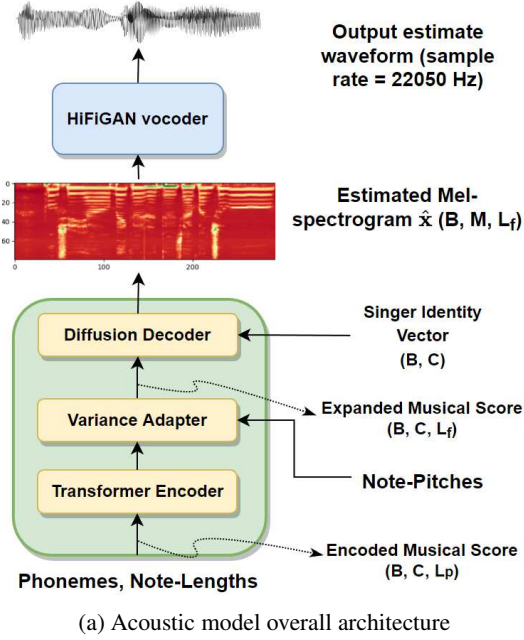}
			\caption{Acoustic model overall architecture}
			\label{fig:overall}
		\end{subfigure}\hfill
		\centering
		\begin{subfigure}[t]{.49\textwidth}
			\centering
			\includegraphics[width=1.0\textwidth]{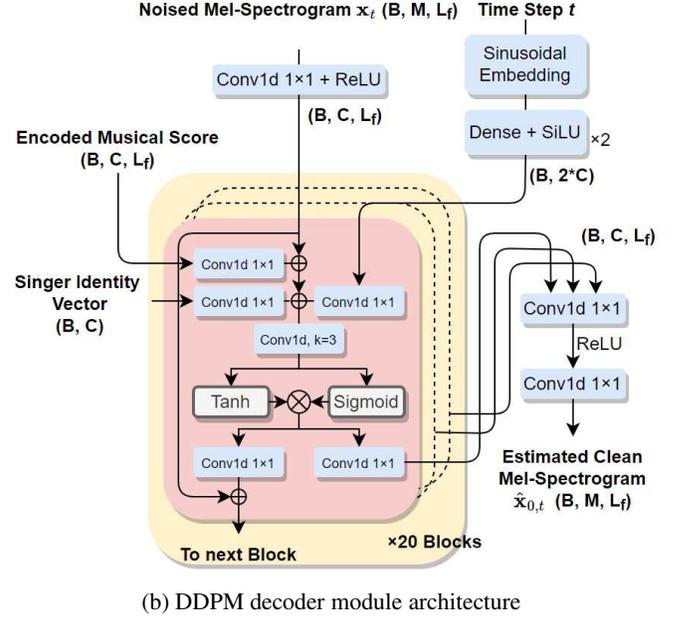}
			\caption{DDPM decoder module architecture}
			\label{fig:ddpm}
		\end{subfigure}
		\caption{Fig ~\ref{fig:overall}: the overall pipeline of the proposed end-to-end SVS system. Fig ~\ref{fig:ddpm}: illustration of the WaveNet blocks used in the diffusion decoder. ($B$: batch size, $C$: number of channels, $L_p$: length of the phone sequence, $L_f$: length of the Mel-spectrogram, $M$: Mel-spectrograms' number of bins.)}
	\end{figure*}
	
	\subsection{Diff-WGAN formulated for SVS} \label{sec:diffwgan-svs}
	The \emph{adversarial objective} demands the conditional discriminator to distinguish the true prior Mel-spectrogram $\mathbf{x}_{t-1}$ and the estimated $\mathbf{\hat{x}}_{t-1} \sim p_{\theta}( \mathbf{\hat{x}}_{t-1}|\mathbf{x}_t )$ generated by the acoustic model according to the definition in Equation ~\ref{est_x_t-1}.
	
	For the SVS task, we formulated the generator and discriminator as:
	\begin{equation} \label{GD}
		\begin{aligned}
			&Generator: G_{\theta}(\mathbf{x}_t, t, \mathbf{ms}, id)\\
			&Discriminator: D_{\phi}(\mathbf{x}_{t-1},\mathbf{x}_t,t,\mathbf{s}_{\mathrm{pho}}, \mathbf{s}_{\mathrm{len}}, \mathbf{s}_{\mathrm{pit}},id)
		\end{aligned}
	\end{equation}
	where $_{\theta}$ and $\phi$ denote the learnable parameters for the two networks, $\mathbf{ms}$ denotes the musical score encoded by the musical score encoder and variance adaptor modules, $id$ denotes singer identity, and $\mathbf{s}_{\mathrm{pho}}, \mathbf{s}_{\mathrm{len}}, \mathbf{s}_{\mathrm{pit}} \in \mathbb{R}^{1\times L_{p}}$ denote the musical score's phone sequence, note-length sequence, and note-pitch sequence.
	
	In addition to the real/fake $\mathbf{x}_{t-1}$/$\mathbf{\hat{x}}_{t-1}$, the discriminator is provided with a real prior $\mathbf{x}_{t}$ and the corresponding time step $t$ as completion of conditions for a noising diffusion transition. In addition, the musical score's information and the singer id $id$ are encoded and fed as auxiliary information since they were found to improve training stability and enhance lingual features in related studies \cite{diffgan_tts, gantts}.
	
	To enforce the \emph{Lipschitz} continuity required by WGAN \cite{wgan}, we applied the \emph{gradient penalty} \cite{wgangp}. Hence we have the minimization criterion for the discriminator as
	\begin{equation} \label{discriminator_loss}
		\begin{aligned}
			\mathcal{L}_\mathrm{D} = \mathcal{L}_{\mathrm{WD}} + \lambda_{GP} \mathcal{L}_{\mathrm{GP}},
		\end{aligned}
	\end{equation}
	where the \emph{Wasserstein distance} criterion is
	\begin{equation}
		\begin{aligned}
			&\mathcal{L}_{\mathrm{WD}} = \sum_{\substack{t \geq 1}} -\mathbb{E}_{q(\mathbf{x}_t)q(\mathbf{x}_{t-1}|\mathbf{x}_t)}\\
			&\quad [ D_{\phi}(\mathbf{x}_{t-1},\mathbf{x}_t,t,\mathbf{s}_{\mathrm{pho}}, \mathbf{s}_{\mathrm{len}}, \mathbf{s}_{\mathrm{pit}},id)] + \\
			& \quad \mathbb{E}_{ q(\mathbf{x}_t) p_{\theta}( \mathbf{\hat{x}}_{t-1}|\mathbf{x}_t ) }
			[D_{\phi}(\mathbf{\hat{x}}_{t-1},\mathbf{x}_t,t,\mathbf{s}_{\mathrm{pho}}, \mathbf{s}_{\mathrm{len}}, \mathbf{s}_{\mathrm{pit}},id]).
		\end{aligned}
	\end{equation}
	with $\lambda_{\mathrm{GP}}=10.0$ for the \emph{gradient penalty}:
	\begin{equation}
		\begin{aligned}
			\mathcal{L}_{\mathrm{GP}} &= 
			\sum_{\substack{t \geq 1}} \mathbb{E}_{\tilde{\mathbf{x}}_{t-1} \sim \mathcal{P}(\tilde{\mathbf{x}}_{t-1})} \\
			&[||\nabla_{\tilde{\mathbf{x}}_{t-1}}D_{\phi}(\mathbf{\hat{x}}_{t-1},\mathbf{x}_t,t,\mathbf{s}_{\mathrm{pho}}, \mathbf{s}_{\mathrm{len}},\mathbf{s}_{\mathrm{pit}},id) ||_{2}-1]^2,
		\end{aligned}
	\end{equation}
	where
	\begin{equation}
		\begin{aligned}
			\tilde{\mathbf{x}}_{t-1} &= \alpha \mathbf{x}_{t-1} + (1-\alpha) \hat{\mathbf{x}}_{t-1}, \alpha  \sim \mathcal{U}(0,1).
		\end{aligned}
	\end{equation}
	
	As defined by \cite{wgan}, the generator's adversarial loss is formulated against that of the discriminator as
	\begin{equation} \label{adv_loss}
		\begin{aligned}
			&\mathcal{L}_{adv} = \sum_{\substack{t \geq 1}} \mathbb{E}_{ q(\mathbf{x}_t)} [D_{\phi}(
			G_{\theta}(\mathbf{x}_t, t, \mathbf{ms}, id)
			,\mathbf{x}_t,t,\\
			&\mathbf{s}_{\mathrm{pho}}, \mathbf{s}_{\mathrm{len}}, \mathbf{s}_{\mathrm{pit}},id)].
		\end{aligned}
	\end{equation}
	
	\section{Proposed System}
	As depicted in Fig.~\ref{fig:overall}, our proposed SVS system is composed of an acoustic model that synthesizes Mel-spectrograms from musical score inputs and a HiFi-GAN neural vocoder \cite{hifigan} that synthesizes waveforms from estimated Mel-spectrograms. The specifications of the acoustic model are presented in Table ~\ref{tab:g_am_spec}.
	
	\subsection{Acoustic Model} \label{acoustic-model}
	
	\subsubsection{\textbf{Transformer Encoder}} \label{sec:encoder}
	The encoder is based on \emph{FastSpeech2} \cite{fastspeech2} and \emph{DiffSinger}'s encoder modules \cite{diffsinger}. Here, we exploit Mandarin phonology to pair a note on the score with a syllable in the lyrics and decompose every syllable into an \emph{initial}-\emph{final} pair, the details of which can be found in our previous work \cite{lstm_mandarin}. Each \emph{initial} and each \emph{final} are treated as a phone in the context of this work. The phone sequence and the note-length sequence are embedded, added together and passed through the Transformer stack to become the phone latent sequence.
	
	\subsubsection{\textbf{Variance Adaptor}} \label{sec:variance-adaptor}
	The \emph{Variance Adaptor} contains a \emph{duration predictor} and a \emph{sequence-length regulator} which expands the length of the phone latent sequence from token-level to frame-level, matching that of the ground truth Mel-spectrograms, as in \cite{fastspeech2, xiaoicesing, diffsinger}. In the meantime, the note-pitch information $\mathbf{s}_{\mathrm{pit}}$ is separately embedded, expanded by the \emph{sequence-length regulator}, and added to the phone latent sequence to produce the frame-level hidden state sequence $\mathbf{ms} \in \mathbb{R}^{C\times L_f}$ that encompasses all the information provided by the musical score.
	
	\subsubsection{\textbf{Diffusion Decoder}} \label{sec:decoder}
	The architecture of the decoder $G_{\theta}(\mathbf{x}_t, t, \mathbf{ms}, id)$ is similar to that of \cite{diffsinger, diffwave}, which is essentially a stacked non-causal conditional WaveNet \cite{wavenet} as illustrated in Fig. ~\ref{fig:ddpm}. With time step condition $t$ encoded through a \emph{sinusoidal embedding} module followed by a stack of two dense layers with \emph{sigmoid linear unit} (SiLU) \cite{silu} non-linearity, the diffusion decoder estimates a clean Mel-spectrogram $\hat{\mathbf{x}}_{0,t}$ from the $t$-step noised input $\mathbf{x}_t$ conditioned on musical score information $\mathbf{ms}$. The estimated clean Mel-spectrogram is then noised through Equation ~\ref{x_t-1} to complete the $p_{\theta}( \mathbf{\hat{x}}_{t-1}|\mathbf{x}_t )$ generation process stated by Equation ~\ref{est_x_t-1}.
	
	\begin{figure*}[t]
		\centering
		\begin{subfigure}[t]{.49\textwidth}
			\centering
			\includegraphics[width=0.95\textwidth]{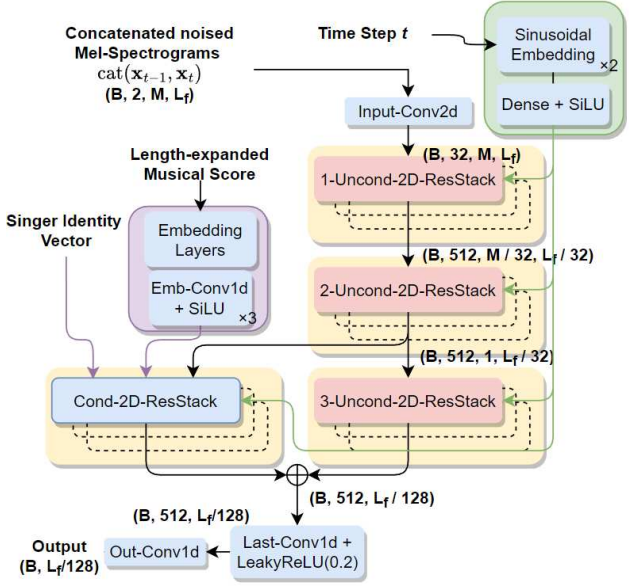}
			\caption{The overall architecture of the \emph{MSC}-discriminator.}
			\label{fig:D_arch}
		\end{subfigure}\hfill
		\centering
		\begin{subfigure}[t]{.49\textwidth}
			\centering
			\includegraphics[width=0.8\textwidth]{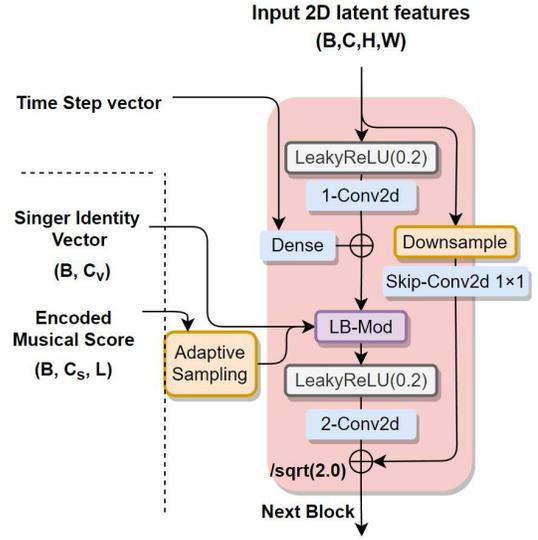}
			\caption{The ResBlock architecture of the \emph{MSC}-discriminator}
			\label{fig:D_resblock}
		\end{subfigure}
		\caption{Fig ~\ref{fig:D_arch}: The overall architecture of the \emph{MSC}-discriminator. Fig ~\ref{fig:D_resblock}: The internal architecture of the \emph{Conditional}-ResBlocks used by the \emph{MSC}-discriminator. The modules to the right of the dashed line are all block-dependent modules/operations. \emph{Unconditional}-ResBlocks takes the same architecture, but no \emph{Singer Identity Vector} and \emph{Musical Score} information is provided and \emph{LB-mod} is not performed. ($H$, $W$: the dimensions of the input downsampled from $M \times L_f$; $C_s$, $C_v$: channel dimensions of the singer identity vector and the encoded musical score respectively. Other symbols are defined the same as in Fig. \ref{fig:overall})}
	\end{figure*}
	
	\begin{table}[t] \label{g_spec}
		\caption{Acoustic Model Specifications}
		\label{tab:g_am_spec}
		\centering
		\begin{scriptsize}		
		\begin{tabularx}{\columnwidth}{c|>{\centering\arraybackslash}m{4cm}}
			\hline
			Module & Hyperparameters
			\\
			\hline
			\textbf{Transformer Decoder} & \begin{tabular}{@{}c@{}}Transformer-layers=4, hidden-dim=256, \\ feed-forward size=1024, kernel-size=9\end{tabular} \\
			\hline
			\textbf{Duration Predictor} & \begin{tabular}{@{}c@{}}CNN-layers=3, hidden-dim=256,\\ kernel-size=3\end{tabular} \\
			\hline
			\textbf{Diffusion Decoder} & \begin{tabular}{@{}c@{}@{}}WaveNet-blocks=20, hidden-dim=256,\\ kernel-size=3, \\sinusoidal-embedding-dim=512\end{tabular} \\
			\hline\hline
			\textbf{Mel-spectrogram} & \begin{tabular}{@{}c@{}}sample-rate=22050Hz, hop-length=256, \\ frame-length=1024, bins=80\end{tabular} \\
			\hline
		\end{tabularx}
	\end{scriptsize}
	\end{table}
	
	\subsection{Musical-Score-Conditioned Discriminator} \label{discriminator}
	The Musical-Score-Conditioned (\emph{MSC})-discriminator $D_{\phi}(\mathbf{x}_{t-1},\mathbf{x}_t,t,\mathbf{s}_{\mathrm{pho}}, \mathbf{s}_{\mathrm{len}}, \mathbf{s}_{\mathrm{pit}},id)$'s backbone is 2D residual convolutional blocks (ResBlock), which are grouped and connected as depicted in Fig. ~\ref{fig:D_arch}.
	
	\subsubsection{\textbf{Residual Block}} \label{sec:msc_d_resblock}
	The \emph{Residual Block} (ResBlock) illustrated in Fig ~\ref{fig:D_resblock} has two variants: the \emph{conditional} ResBlock and the \emph{unconditional} ResBlock. To adapt to the varying level of noise, a time step vector is projected through a dense layer and added to the residual latent features. For conditional blocks, the singer identity vector and the adaptive-interpolated encoded musical score sequence condition the residual latent features through the \emph{Locally-Biased Modulation} (LB-mod) algorithm.
	
	\subsubsection{\textbf{\emph{Locally-Biased Modulation}}} \label{sec:msc_d_LBMod}
	The \emph{Locally-Biased Modulation} (LB-mod) algorithm in the \emph{conditional} ResBlock was inspired by \cite{stylegan} but differs from it in preserving the temporal dimension. \emph{LB-mod} takes the adaptive-interpolated encoded musical score $\mathrm{\mathbf{ms}} \in \mathbb{R}^{C_s\times W}$, the singer identity vector $\mathrm{\mathbf{v}} \in \mathbb{R}^{C_v\times 1}$, and the latent features $\mathrm{\mathbf{y}} \in \mathbb{R}^{C\times H\times W}$.
	
	With two dense layers, LB-mod first projects and combines $\mathrm{\mathbf{ms}}$ and $\mathrm{\mathbf{v}}$ to produce the condition sequence $\mathrm{\mathbf{s}} \in \mathbb{R}^{2C\times W}$ as in Equation ~\ref{lb_mod_proj}:
	\begin{equation} \label{lb_mod_proj}
		\begin{aligned}
			\mathrm{\mathbf{s}}^{2C\times W} =
			\mathrm{Dense}^{C_s \to 2C}_\mathrm{ms}(\mathrm{\mathbf{ms}}) + 
			\mathrm{Dense}^{C_v \to 2C}_\mathrm{id}(\mathrm{\mathbf{v}})
		\end{aligned}
	\end{equation}
	
	Subsequently, the modulation is performed as a linear-transform-and-bias operation defined in Equation ~\ref{lb_mod_mod}:
	\begin{equation} \label{lb_mod_mod}
		\begin{aligned}
			&\mathbf{LB\mbox{-}Mod}(\mathrm{\mathbf{y}}^{C\times H\times W}, \mathrm{\mathbf{s'}}^{2C\times \mathrm{H} \times W}):=\\
			& \quad \mathrm{\mathbf{s'}}_{1:C} \odot \mathrm{\mathbf{y}} + \mathrm{\mathbf{s'}}_{C+1:2C},
		\end{aligned}
	\end{equation}
	where $\mathrm{\mathbf{s'}}^{2C\times \mathrm{H} \times W}$ is $\mathrm{\mathbf{s}}^{2C \times W}$ reshaped by expanding and repeating along the second dimension.
	
	\subsection{Training Objective} \label{loss}
	The training objective of the acoustic model generator contains three elements: the duration loss, the reconstruction loss, and the adversarial loss.
	
	In the training process, the ground truth phone durations $\mathbf{D}^{1\times L_{p}}$ are used, while the estimated durations $\hat{\mathbf{D}}^{1\times L_{p}}$ are used in the inference stage. Therefore, the \emph{variance adaptor} returns both the encoded musical score information and the estimated duration sequence, for which a mean-square-error (MSE) is taken against the ground truth duration sequence:
	\begin{equation} \label{dur_loss}
		\begin{aligned}
			\mathcal{L}_\mathrm{dur} = \parallel \mathbf{D} - \hat{\mathbf{D}} \parallel_2.
		\end{aligned}
	\end{equation}
	
	The diffusion decoder's Mel-spectrogram reconstruction loss is calculated as the L1-distance between $\mathbf{x}_0$ and $\mathbf{\hat{x}}_{0,t}$:
	\begin{equation} \label{recon_loss}
		\begin{aligned}
			\mathcal{L}_\mathrm{recon} = \parallel \mathbf{x}_0 - \mathbf{\hat{x}}_{0,t} \parallel_1.
		\end{aligned}
	\end{equation}
	
	With the adversarial loss defined by Equation ~\ref{adv_loss}, the acoustic model generator parameters $\theta$ are updated with gradients calculated with the weighted sum of the three objectives:
	\begin{equation} \label{train_obj}
		\begin{aligned}
			\mathcal{L}_G &= \mathcal{L}_\mathrm{dur} + \lambda_{\mathrm{recon}} \mathcal{L}_\mathrm{recon} + \lambda_{\mathrm{adv}} \mathcal{L}_\mathrm{adv}.
		\end{aligned}
	\end{equation}
	
	The weight for the reconstruction loss $\lambda_{\mathrm{recon}}$ and the adversarial loss $\lambda_{\mathrm{adv}}$ were both set to 1.0 for the default mixed setup. For the WGAN-only setup in the later experiment section, $\lambda_{\mathrm{recon}}$ was set to 0.
	
	\subsection{HiFi-GAN Neural Vocoder} \label{vocoder}
	As the backend audio synthesizer for our SVS system, HiFi-GAN \cite{hifigan} \footnote{\url{https://github.com/jik876/hifi-gan}} was chosen for its compute efficiency, state-of-the-art synthesis audio quality, and robustness shown across multiple related works \cite{fastspeech2, hifisinger, diffsinger, diffgan_tts}. For our implementation and experiments, HiFi-GAN v2 was adopted with the number of initial channels modified to 256.
	
	After HiFi-GAN was pre-trained on the Mpop600 dataset, it was fine-tuned to mitigate the discrepancies between real and synthesized Mel-spectrograms. In the last 35k steps of acoustic model training, the pre-trained HiFi-GAN vocoder was loaded and trained with the estimated Mel-spectrograms of the Diff-WGAN acoustic model stochastically mixed into their training data according to the probability density function defined in Equation ~\ref{replace_pr}:
	\begin{equation} \label{replace_pr}
		\begin{aligned}
			\mathrm{Pr}[\text{replace } \mathbf{x}_0 \text{ with } \mathbf{\hat{x}}_{0,t}] = (1-\frac{t-1}{T})^2.
		\end{aligned}
	\end{equation}
	
	\section{Experiments}
	\subsection{Dataset} \label{dataset}
	All the experiments in this work were conducted with our lab's Mpop600 Mandarin singing voice dataset \cite{mpop600}. The Mpop600 dataset contains the singing voices of two female and two male singers. Each singer contributed 3 hours of singing voice audio consisting of 150 different song excerpts. In total, the entire Mpop600 dataset is 12-hour-long with 600 song excerpts. All the songs in this dataset were pop-music sung in Mandarin and recorded without background instruments. The audio and phone-transcribed lyrics were forced-aligned to frame-level by the open-source Speech-Aligner\footnote{\url{https://github.com/open-speech/speech-aligner}}, the details of which can be found in our previous work \cite{mpop600, lstm_mandarin}. For all the experiments, the audio recordings were down-sampled from 96 kHz, 24-bits to 22.05 kHz, 16-bits.
	
	For evaluation, two song excerpts from each singer's 150 song excerpts were reserved and not seen by the model in the training process. To train the models with mini-batches, every song excerpt was segmented into smaller audio samples with audio durations of 6 to 12 seconds.
	
	\subsection{Training Setup} \label{training}
	The experiments were conducted with audio samples produced through the following setups:
	
	\begin{itemize}
		\item \rm{\textbf{Reference}}: ground truth Mpop600 audio samples sung by human singers down-sampled to 22.05 kHz.
		\item \rm{\textbf{Resynthesized}}: ground truth audio samples converted to Mel-spectrograms and re-synthesized by the HiFi-GAN vocoder at 22.05 kHz sample rate.
		\item \rm{\textbf{model-FFT}}: an end-to-end SVS system with a \emph{Feed-forward Transformer} backbone as the acoustic model decoder trained on an L1 Mel-spectrogram estimation loss.
		\item \rm{\textbf{model-Diff-L1}} ($T=4$): the proposed system with the DDPM acoustic model trained exclusively on the L1 Mel-spectrogram reconstruction loss without WGAN.
		\item \rm{\textbf{model-Diff-Mixed}} ($T=4$): the proposed system with the DDPM acoustic model trained with both the L1 Mel-spectrogram reconstruction loss and the WGAN adversarial loss.
		\item \rm{\textbf{model-Diff-WGAN}} ($T=4$): the proposed system with the DDPM acoustic model trained exclusively by WGAN without the L1 Mel-spectrogram reconstruction loss.
	\end{itemize}
	
	All the end-to-end synthesis models utilized the HiFi-GAN vocoder for waveform generation and Mel-specrtograms as the acoustic features. The experimented acoustic models were trained by an \emph{AdamW} \cite{adamw} optimizer with L2 weight decay of $10^{-6}$ while the the HiFi-GAN vocoder was pre-trained with an \emph{Adam} \cite{adam} optimizer without weight decay.
	
	For model-FFT, \emph{AdamW} hyperparameters $\beta_{1}=0.9$, $\beta_{2}=0.98$, $\mathrm{lr} = 10^{- 4}$ were applied with the learning rate scheduling proposed for \emph{Transformer}s in \cite{attention}. For all the diffusion-based models, the \emph{AdamW} hyperparameters were $\beta_{1}=0.5$, $\beta_{2}=0.9$. Since model-Diff-L1 was trained solely on reconstruction objectives, its learning rate was exponentially decayed by a rate of 0.999 at the start of every epoch with an initial learning rate $\mathrm{lr}=10^{-4}$. For the two Diff-WGAN acoustic models (model-Diff-Mixed, model-Diff-WGAN), the learning rate was $10^{-4}$ for both the generator and the discriminator, and no learning rate scheduling was applied.
	
	All the acoustic models were trained to 210k steps, taking 16 hours for model-FFT, 20 hours for model-Diff-L1, and 48 hours for the two Diff-WGAN models since the discriminator and the acoustic model were updated with a $2:1$ ratio as WGAN requires the discriminator to converge faster than the generator \cite{wgan,wgangp}. These training sessions were completed on one RTX 3090 graphics card with a batch size of 8.
	
	\subsection{Objective Evaluation} \label{sec:eval_obj}
	For perceptual similarity, we employed two objective metrics: \emph{Mel-Cepstral-Distortion} (MCD) commonly adopted for audio signals and the \emph{Multi-Scale Mean Structural Similarity} (MS-SSIM) \cite{mssim} metric. Here, MS-SSIM was applied to the Mel-spectrograms of the synthesized and ground truth signal to evaluate their structural similarity on the spectral domain as images \cite{freq_mssim}.
	
	To evaluate the fundamental frequency (F0) fidelity of the SVS system on the test dataset, we employed the \emph{root-mean-square-error} (RMSE) and \emph{Pearson-correlation} (Corr) metrics on the F0 sequences extracted by the \emph{WORLD} vocoder \cite{world} \footnote{\url{https://github.com/JeremyCCHsu/Python-Wrapper-for-World-Vocoder}}. To eliminate the effects of rhythmic swings, \emph{Dynamic Time Warping} (DTW) was first performed between the ground truth signal and the compared signal with an 
	open-source fast-DTW tool \footnote{\url{https://github.com/slaypni/fastdtw}}. For the musical context of this work, the F0-RMSE values were calculated on the \emph{semitone} scale.
	
	The objective evaluation results of the compared systems are presented in Table ~\ref{tab:obj_results}.
	
	\begin{table}[t]
		\begin{center}
			\begin{scriptsize}			
			\begin{tabularx}{\columnwidth}{c|cccc}
				\hline
				System&MS-SSIM ${}^\uparrow$&MCD ${}^\downarrow$&F0-RMSE ${}^\downarrow$&F0-Corr ${}^\uparrow$ \\
				\hline\hline
				\textbf{Resynthesized}&0.985&2.613&0.273&0.936 \\ 
				\hline
				\textbf{FFT}&0.756&8.206&0.628&0.751 \\
				\hline
				\textbf{Diff-L1}&\textbf{0.899}&\textbf{5.702}&\textbf{0.411}&\textbf{0.882}\\ 
				\hline
				\textbf{Diff-Mixed}&0.879&6.348&0.463&0.881\\ 
				\hline
				\textbf{Diff-WGAN}&0.886&6.130&0.438&0.878\\
				\hline
			\end{tabularx}
		\end{scriptsize}
		\end{center}
		\caption{Objective evaluation results of different systems.}
		\label{tab:obj_results}
	\end{table}
	
	\subsection{Mean-Opinion-Score evaluation} \label{sec:eval_subj}
	To understand each system's synthesis quality perceived by the human listener, 10 participants were asked to rate each sample out of $\{1,2,3,4,5\}$, where \emph{1} indicates \emph{worst} and \emph{5} indicates \emph{best} in terms of \emph{"naturalness concerning human singers"}. Two samples from each of the four singers were used in the MOS test. The MOS data are presented in Table ~\ref{tab:mos_results} with their $95\%$ confidence intervals.
	
	\begin{table}[t]
		\begin{center}
			\renewcommand{\arraystretch}{1.5}
			\centering
			\begin{tabular}{|c|c |}
				\hline
				System & 
				MOS Score $(1\sim 5)$ \\
				\hline\hline
				\textbf{Reference} & $4.76 \pm 0.099$\\ 
				\hline
				\textbf{Re-synthesized} & $4.15 \pm 0.243$ \\ 
				\hline \hline
				\textbf{model-FFT} & $2.55 \pm 0.447$ \\
				\hline
				\textbf{model-Diff-L1} & $\mathbf{3.79 \pm 0.242}$ \\ 
				\hline
				\textbf{model-Diff-Mixed} & $3.49 \pm 0.481$ \\ 
				\hline
				\textbf{model-Diff-WGAN} & $2.88 \pm 0.279$ \\
				\hline
			\end{tabular}
		\end{center}
		\caption{Mean Opinion Score (MOS) evaluation results of different systems collected on 10 human participants.}
		\label{tab:mos_results}
	\end{table}
	
	\subsection{Preference Test on Musical Expressiveness} \label{sec:music_pref}
\begin{figure}[htbp]
		\centering
		\includegraphics[width=0.95\linewidth]{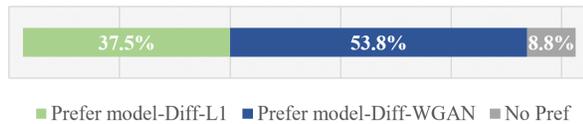}
		\caption{Musical expressiveness preference test results on 20 participants.}
		\label{fig:music_pref}
	\end{figure}

	To particularly examine the proposed Diff-WGAN architecture's effects on musical expressiveness, model-Diff-L1 and model-Diff-WGAN were subjected to a preference test with 20 participants. Four extra audio segments were sampled from the validation set of the two skilled singers (one male and one female) who exhibit more complex variations musically. The audio samples were formulated into four questions. In each question, the human participants had to choose the sample they deemed \emph{more musically expressive disregarding pronunciation accuracy or audio quality} or choose the \emph{No Preference} option if they deemed the two on par. The result of the preference test is shown in Fig ~\ref{fig:music_pref}.
	
	\section{Discussion and Analysis} \label{sec:disc_analysis}
	\subsection{Discussion on Quantitative Evaluation Results} \label{sec:eval_results}
	In Table ~\ref{tab:mos_results}'s MOS test results, model-Diff-L1 received the highest score. Looking at the measurements presented in Table ~\ref{tab:obj_results}, model-Diff-L1's advantage can likely be attributed to its reconstructive L1 objective. Since its target was only to best-estimate the ground truth Mel-spectrograms in the L1 distance, model-Diff-L1 naturally had a low overall distortion to the ground truth, as evident from the objective measurements. This attribute means model-Diff-L1 robustly produces the high-energy spectral features that mimic those produced by human singers and rids the synthesized Mel-spectrograms of significant artifacts that greatly impact the perceived quality of an audio segment in the singing voice context.
	
	In comparison, model-Diff-Mixed and model-Diff-WGAN's objectives were not to estimate the ground truth Mel-spectrograms but their generation distributions; therefore, the generator was enforced to also reproduce the variations and inconsistencies of human singers. However, in the context of sung music, human listeners give audio samples with any artifacts or errors in pitch or pronunciation significantly lower scores regardless of the rest of their properties. Hence, the less consistent Diff-WGAN-based models received lower scores in both the objective measurements and the MOS test.
	
	Nevertheless, model-Diff-WGAN crucially beats model-Diff-L1 in the musical expressiveness preference test. While the best strategy for model-Diff-L1 to optimize for reconstruction objectives was to over-smooth these variations, model-Diff-WGAN optimized towards the data distribution estimated by the acoustic model discriminator, which injects variations and details not just acoustically but also on meta-features as musical expression. This observation is further investigated in the following qualitative study.

	\subsection{Qualitative Study on Generated Mel-Spectrograms} \label{sec:eval_melspec}
	
	\begin{figure}[htbp]
		\centering
		\includegraphics[width=0.9\linewidth]{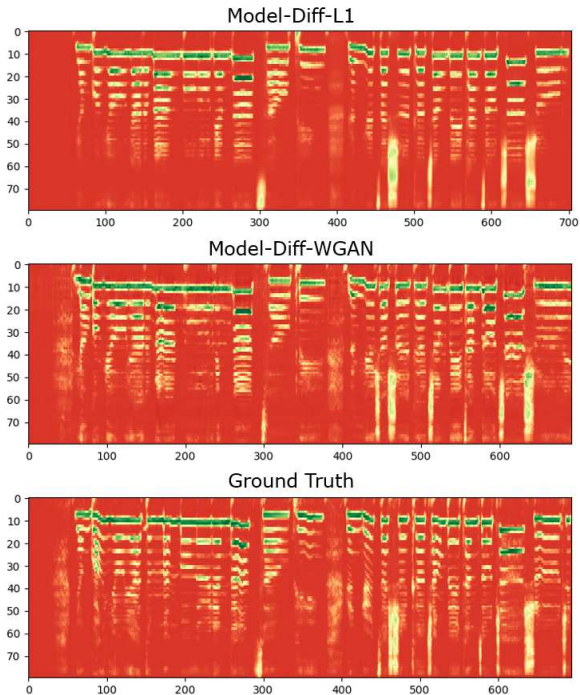}
		\caption{A set of Mel-spectrogram samples from model-Diff-L1, model-Diff-WGAN, and extracted directly from ground truth human singer's audio in the validation dataset.}
		\label{fig:mel_spec_0}
	\end{figure}
	
	The synthesis characteristics of acoustic models with or without WGAN can be understood by observing their resulting Mel-spectrograms. As illustrated in Fig. ~\ref{fig:mel_spec_0}, model-Diff-L1 trained only with reconstruction loss and model-Diff-WGAN trained only with WGAN produced Mel-spectrograms with distinct styles. Model-Diff-WGAN's synthesized Mel-spectrogram contains high-frequency components, noises during silent segments, and variations of energy reminiscent of the ground truth Mel-spectrogram.
	
	Contrarily, the high-frequency details on the model-Diff-L1's spectrogram are less prominent and those present are also less granular than the high-frequency details on model-Diff-WGAN's or the ground truth's Mel-spectrogram. Nevertheless, when we focus on the potent high-energy components in the lower frequency bins, model-Diff-L1's Mel-spectrogram shows a more consistent and smooth change in energy, and the fundamental frequency contour exhibits clear-cut discontinuities in inter-syllable unvoiced consonants. These properties combined mean the resulting signal does not have unnatural artifacts and gives the perception of properly accentuated syllable pronunciations.
	
	Based on these observations, it can be concluded that the WGAN has indeed prevented the acoustic model from over-smoothing, but the variations it introduced also made the acoustic model less consistent and prone to error. As human listeners are intolerant to errors in music, further investigations have to be made to leverage the correctness enforced by the reconstruction objective and the
	spectral details and expressiveness brought about by the WGAN.

	\subsection{Remark on Training Stability and Convergence} \label{sec:convergence}
	One of the key advantages of the proposed combination of DDPM and WGAN is its stability. In the training process, the discriminator loss of the Diff-WGAN acoustic models \emph{Wasserstein} decreased monotonically, and no mode collapse has been encountered in the experimented configurations even for model-Diff-WGAN, in which no reconstruction loss was employed as constraint or guidance.
	
	\section{Conclusions}
	This work proposed an acoustic model based on a combined architecture of diffusion denoising probabilistic model (DDPM) and \emph{Wasserstein} generative adversarial network (WGAN). To exploit the singing voice synthesis (SVS) formulation, the discriminator was designed to be conditioned not only on the synthesis features but also on the musical score and singer identity. This proposed acoustic model architecture was implemented with an integrated HiFi-GAN vocoder to form a multi-singer end-to-end singing voice synthesis system trained on the Mpop600 Mandarin singing voice dataset. Although the proposed architecture was not fully exploited to comprehensively outperform the baseline diffusion model, the preference test and the qualitative study suggested that the addition of WGAN fulfilled its purpose of prompting musical expressiveness and enforcing high-frequency acoustic details. Moreover, this adversarial acoustic model was shown to converge without the reconstruction objective as guidance, thereby proving the convergence and stability of the proposed Diff-WGAN architecture.
	
	\section{Acknowledgment}
	This research is supported by the Ministry of Science and Technology of Taiwan under Grant No. 109-2221-E-007-094-MY3.
	
	\bibliographystyle{ieeetr}
	\bibliography{refs}
	
\end{document}